\documentclass{article}
\usepackage{graphicx}
\usepackage{amsmath}
\usepackage{amsfonts}
\usepackage{amssymb}

\begin{document}

\title{Restart Strategies and Internet Congestion}
\author{Sebastian M. Maurer and Bernardo A. Huberman\\Xerox Palo Alto Research Center, Palo Alto, CA\ 94304}
\maketitle
\begin{abstract}
We recently presented a methodology for quantitatively reducing the risk and
cost of executing electronic transactions in a bursty network environment such
as the Internet. In the language of portfolio theory, time to complete a
transaction and its variance replace the expected return and risk associated
with a security, whereas restart times replace combinations of securities.
While such a strategy works well with single users, the question remains as to
its usefulness when used by many. By using mean field arguments and
agent-based simulations, we determine that a restart strategy remains
advantageous even if everybody uses it.\newpage
\end{abstract}

\section{Introduction}

The impressive growth in the number of Internet users (from 61 million in 1996
to over 150 million today) has led to a radical increase in the volume of data
that is transmitted at any given time. Whereas a few years ago email was the
preponderant component of Internet traffic, the Web, with its rich and varied
content of images and text, makes up for most of the transmitted data today.
In addition, financial and other forms of electronic transactions put a
premium on mechanisms that ensure timely and reliable transactions in
cyberspace. This is an important problem given the bursty nature of Internet
congestion \cite{congestion}, which leads to a large variability in the risk
and cost of executing transactions.

Earlier, we presented a methodology for quantitatively managing the risk and
cost of executing transactions in a distributed network environment
\cite{lukose}. By associating cost with the time it takes to complete the
transaction, and risk with the variance in that time, we considered different
methods that are analogous to asset diversification, and which yield mixed
strategies that allow an efficient trade-off between the average and the
variance in the time a transaction will take. Just as in the case of financial
portfolios, we found that some of these mixed strategies can execute
transactions faster on average and with a smaller variance in their speed.

A potential problem with this portfolio methodology is that if everybody uses
it, the latency characteristics of the Internet might shift so as to render
the method useless. It this were the case, one would be confronted with a
classical social dilemma, in which cooperation would amount to restraining
from the use of such strategies and defection on their exploitation
\cite{dilemmas}\cite{dynamics}. Thus it becomes important to determine the
effects that many users employing a portfolio strategy have on Internet
latencies and their variances. Equally relevant is to determine what the
increased latencies are as a function of the fraction of users deciding to
employ such strategies.

In order to investigate this issue, we conducted a series of computer
simulations of a group of agents deciding asynchronously whether to use the
Internet or not. The agents base their decision on knowledge of the congestion
statistics over a past window of time. We found that when every agent uses the
portfolio strategy there is still a range of parameters such that a) a
portfolio exists and b) all agents are better off using it than not. Even when
all agents do so, the optimum restart strategy leads to a situation no worse
than when no one uses the restart strategy.

\section{Restart Strategies}

Anyone who has browsed the World Wide Web has probably discovered the
following strategy: whenever a web page takes too long to appear, it is useful
to press the reload button. Very often, the web page then appears instantly.
This motivates the implementation of a similar but automated strategy for the
frequent ''web crawls'' that many Internet search engines depend on. In order
to ensure up-to-date indexes, it is important to perform these crawls quickly.
More generally, from an electronic commerce perspective, it is also very
valuable to optimize the speed and variance in the speed of transactions,
automated or not, especially when the cost of performing those transactions is
taken into account. Again, restart strategies may provide measurable benefits
for the user.

\begin{figure}[tbh]
\begin{center}
\includegraphics[scale=0.5]{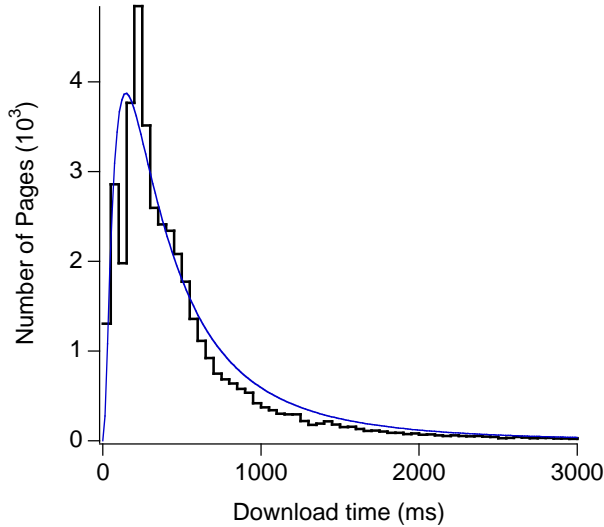}
\end{center}
\caption{Download time in milliseconds of the index.html file on the main page
of over forty thousand web sites, with a fit to a log-normal distribution. The
parameters are $\sigma= 0.99$ and $\mu= 5.97$.}%
\end{figure}

The histogram in Figure 1 shows the variance associated with the download time
for the text on the main page of over 40,000 web sites. Based on such
observations, Lukose and Huberman \cite{lukose} recently proposed an economics
based strategy for quantitatively managing the risk and cost of executing
transactions on a network. By associating the cost with the time it takes to
complete the transaction and the risk with the variance in that time, they
exploited an analogy with the modern theory of financial portfolio management
first suggested in a more general context in a previous paper
\cite{Huberman1997}. In modern portfolio theory, the fact that investors are
risk-averse means that they may prefer to hold assets from which they expect a
lower return if they are compensated for the lower return with a lower level
of risk exposure. Furthermore, it is a non-trivial result of portfolio theory
that simple diversification can yield portfolios of assets which have higher
expected return as well as lower risk. In the case of latencies on the
Internet, thinking of different restart strategies is analogous to asset
diversification: there is an efficient trade-off between the average time a
request will take and the variance or risk in that time.

Consider a situation in which a message has been sent and no acknowledgment
has been received for some time. This time can be very long in cases where the
latency distribution has a long tail. One is then faced with the choice to
either continue to wait for the acknowledgment, to send out another message
or, if the network protocols allow, to cancel the original message before
sending out another. For simplicity, we consider the case in which it is
possible to cancel the original message before sending out another at the
time. $P(t)$, the probability that a page has been successfully downloaded in
time less than t, is given by%

\begin{align}
P(t) = p(t)\,\,\,\,\,\,\,\,\mathrm{if}\,\,\,\,t \le\tau\,,\nonumber\\
P(t) = (1 - \int_{0}^{\tau}p(t) dt) P(t - \tau)\,\,\,\,\,\,\,\,\mathrm{if}%
\,\,\,\,t > \tau\,\,.\nonumber
\end{align}

\noindent where $p(t)$ is the probability distribution for the download time
without restart. The latency and risk in loading a page is then given by%

\begin{align}
\langle t \rangle= \int_{0}^{\infty}t P(t) dt\,,\nonumber\\
\sigma= \sqrt{Var(t)} = \sqrt{\langle(t - \langle t \rangle)^{2} \rangle
}\,\,.\nonumber
\end{align}

If we allow an infinite number of restarts, the recurrence relation above can
be solved in terms of the partial moments $M_{n}(\tau) = \int_{0}^{\tau} t^{n}
P(t) dt$:%

\begin{align}
\langle t \rangle= \frac{1}{M_{0}} (M_{1} + \tau(1 - M_{0})) \,,\nonumber\\
\langle t^{2} \rangle= \frac{1}{M_{0}} (M_{2} + \tau(1 - M_{0}) (2 \frac
{M_{1}}{M_{0}} + \tau(\frac{2}{M_{0}} - 1))) \,.\nonumber
\end{align}

In the case of a log-normal distribution $p(t) = \frac{1}{\sqrt{2 \pi} x
\sigma} \exp(- \frac{{(\log x -\mu)^{2}}}{2\,\sigma^{2}})$, $\langle t
\rangle$ and $\langle t^{2} \rangle$ can be expressed in terms of the error function:%

\begin{align}
M_{n}(\tau) = \frac{1}{2} \exp(\frac{\sigma^{2} n^{2}}{2} + \mu n) (1 +
\mathrm{{erf}(\frac{\log\tau- \mu}{\sigma\sqrt{2}}- \frac{\sigma n}{\sqrt{2}%
})) .}\nonumber
\end{align}

The resulting $\langle t\rangle$ versus $\sigma$ curve is shown in Fig. 2(a).
As can be seen, the portfolio has a cusp point that represents the restart
time $\tau$ that is preferable to all others. No strategy exists in this case
with a lower expected waiting time (at possibly the cost of a higher risk) or
with a lower risk (at possibly the cost of a higher expected waiting time).
The location of the cusp can be translated into the optimum value of the
restart time to be used to reload the page.

\begin{figure}[ptb]
\begin{center}
\includegraphics[scale=0.6]{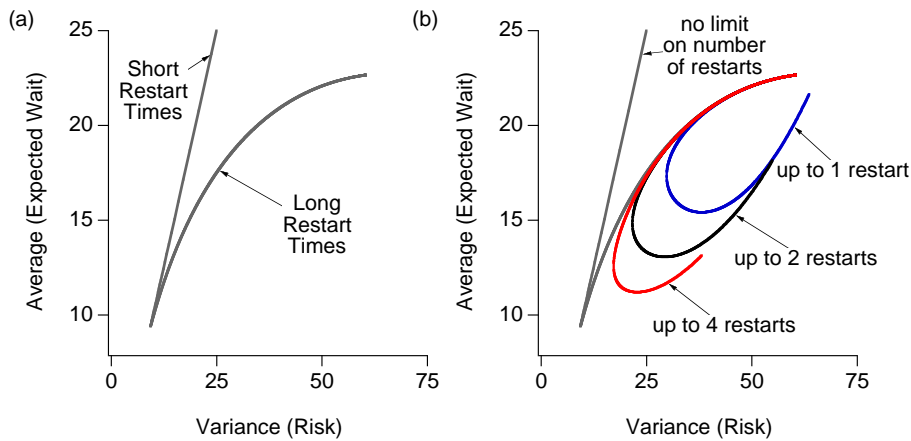}
\end{center}
\caption{(a) Expected average latency versus risk (variance) calculated for a
log-normal distribution with $\mu= 2$ and $\sigma= 1.5$. The curve is
parametrized over a range of restart times. (b) Family of curves obtained when
we limit the maximum number of allowed restarts.}%
\end{figure}

There are many variations to the restart strategy described above. In
particular, in Fig. 2(b), we show the family of curves obtained from the same
distribution used in (a), but with a restriction on the maximum number of
restarts allowed in each transaction. Even a few restarts yield an improvement.

Clearly, in a network without any kind of usage-based pricing, sending many
messages to begin with would be the best strategy as long as we do not
overwhelm the target computer. On the other hand, everyone can reason in
exactly the same way, resulting in congestion levels that would render the
network useless. This paradoxical situation, sometimes called a social
dilemma, arises often in the consideration of ''public goods'' such as natural
resources and the provision of services which require voluntary cooperation
\cite{hardin}. This explains much of the current interest in determining the
details of an appropriate pricing scheme for the Internet, since users do
consume Internet bandwidth greedily when downloading large multimedia files
for example, without consideration of the congestion caused by such activity.

Note that the histogram in Figure 1 represents the variance in the download
time between different sites, whereas a successful restart strategy depends on
a variance in the download times for the same document on the same site. For
this reason, we can not use the histogram in Figure 1 to predict the
effectiveness of the restart strategy. While a spread in the download times of
pages from different sites reduces the gains that can be made using a common
restart strategy, it is possible to take advantage of geography and the time
of day to fine tune and improve the strategy's performance. As a last resort,
it is possible to fine tune the restart strategy on a site per site basis.

As a final caution, we point out that with current client-server
implementations, multiple restarts will be detrimental and very inefficient
since every duplicated request will enter the server's queue and will be
processed separately until the server realizes that the client is not
listening to the reply. This is an important issue for a practical
implementation, and we neglect it here: our main assumption will be that the
restart strategy only affects the congestion by modifying the perceived
latencies. This will only be true if the restart strategy is implemented in an
efficient and coordinated way on both the client and server side.

\section{Restart Strategies for Multiple Users}

While restart strategies based on a portfolio approach may improve returns for
a single user, it is important to consider whether the use of such strategies
leads to a different dilemma. What happens when every user makes use of the
restart strategy? This question is analogous to the problem of adjustments to
equilibrium encountered in finance \cite{Sharpe1970}.

In order to investigate the effects on congestion of many agents using a
restart strategy, we performed a series of computer simulations of a group of
agents deciding asynchronously whether to use the Internet or not. The agents
base their decision on knowledge of the congestion statistics over a past
window of time. As shown by Huberman and Glance \cite{beliefs}, such a
collective action dilemma leads to an optimal strategy which is basically
determined by a threshold function: cooperate if parameters of the problem are
such that a critical function exceeds a certain value and defect otherwise. In
terms of the dilemma posed by using the Internet \cite{congestion} this
translates into downloading pages if latencies are below a certain value and
not doing so if they exceed a particular value. Thus, each agent is restricted
to a simple binary decision, and the dynamics are those of a simple threshold
model with uncertainties built in.

With this in mind, we model each agent as follows: an agent measures the
current congestion, expressed in arbitrary ''latency time'' units. Since the
latency is a function of the number of users, and their number fluctuates in
time, it is reasonable to make the decision to use or not to use the restart
strategy a function of the histogram of the load over a past window of time.
This histogram is used to calculate the perceived latency time using different
strategies, as described in the previous section. The agent compares the
perceived latency to a threshold: if the former is larger, he decides to
''cooperate'' and refrains from using the Internet. If the latency is short
enough, he decides to ''defect''. Agents make these decisions in an
asynchronous fashion, with exponentially distributed waiting times \cite{pnas}.

We assume that the load created by an agent who decides to make use of the
network's resources does not depend on whether or not he uses the restart
strategy. This is only true if the server can efficiently detect multiple
requests and cancel the superfluous ones, in order to avoid sending the same
data to the client multiple times. Current implementations do not offer this
feature. As a result, while a restart strategy may be beneficial to a single
user, the net effect will be to cause more congestion.

In order to calculate the latency $\lambda$ as a function of the number of
users $N_{D}$, we use the average waiting time for a M/M/1 queue
\cite{MM1queue}, with a capacity one larger than the total number of agents
$N$:%

\begin{align}
\lambda= 1 / (1 + N - N_{D}).\nonumber
\end{align}

Note that this simple M/M/1 queue model is not meant to provide anything more
than an intuitive justification for the value of the latency and the
qualitative behavior of its fluctuations (especially their correlations). In
particular, the notion of a single M/M/1 queue is inconsistent with the idea
of a restart strategy.

As it stands, this model would simply relax to an equilibrium in which the
number of users is such that the latency is the threshold latency. To remedy
this, we add fluctuations in the latency times using multiplicative noise
(taken from a gaussian distribution with unit mean). This multiplicative noise
will also be correlated in time (we model it as an Ornstein-Uhlenbeck
process). This correlation is a crucial aspect of the model: agents who arrive
at the same time should experience a similar congestion. If the noise were
completely uncorrelated, agents might as well be on different networks. (We
also performed simulations using additive noise. This produced no qualitative
changes in the results).

A summary of the model is given in the appendix. It includes a list of all the
parameters and the pseudocode for the simulation.

Since the restart strategy is designed to reduce the effect of congestion, we
expect that as more users employ it the congestion will increase. However,
this will be acceptable as long as the perceived latency \textit{with}
restarts does not become worse than the latency originally was without
restarts. This should be the case, as the following mean field model indicates.

In a mean field model, $f_{d}$, the average fraction of agents who defect, is
determined by the differential equation%

\begin{align}
\frac{df_{d}}{dt} = - \alpha(f_{d} - \rho(f_{d})),\nonumber
\end{align}

\noindent where in the context of this paper $\rho(f_{d})$ is the average
probability that the perceived latency is below threshold (i.e. the
probability that a cooperating agent will decide to defect \cite{Huberman1988}%
), and $\alpha$ is the frequency with which agents evaluate their decisions to
cooperate or not. The inverse of this frequency sets the time scale for all
processes in the simulations. In the presence of imperfect knowledge modeled
by a gaussian distribution (with width $\sigma$) of perceived utilities,%

\begin{align}
\rho(f_{d}) = \frac{1}{2}(1 + \mathrm{{erf}(\frac{U(f_{d}) - U_{c}}%
{\sigma\sqrt{2}}))}\nonumber
\end{align}

\noindent where $U(f_{d})$ is the utility of defecting given that a fraction
$f_{d}$ of agents is defecting, and $U_{c}$ is the threshold utility below
while users will cooperate. Expanding around values of $f_{d}$ such that
$U(f_{d}) = U_{c}$, and setting the right hand side of the differential
equation equal to $0$ in order to find the equilibrium point, we obtain%

\begin{align}
U(f_{equ}) = U_{equ} = U_{c} + \sqrt{2} \sigma(2 f_{equ} - 1).\nonumber
\end{align}

\begin{figure}[tbh]
\begin{center}
\includegraphics[scale=0.5]{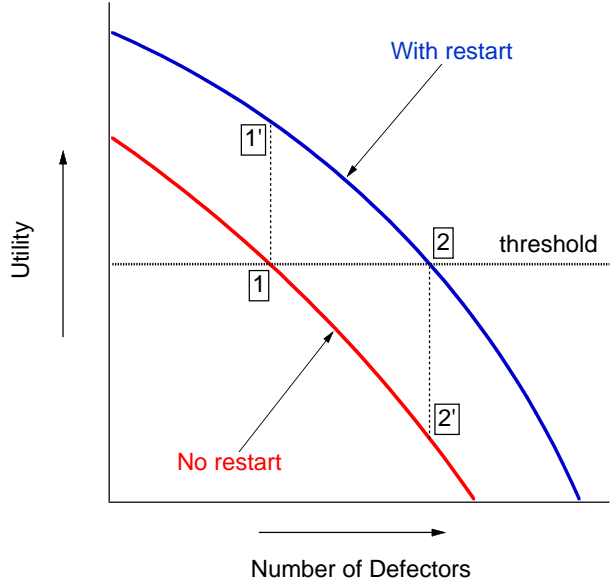}
\end{center}
\caption{Qualitative expectations for the model. See text for description.}%
\end{figure}

Thus, whatever the details of the threshold based multi-agent model, an
average equilibrium number of defectors can be mapped onto an expected utility
(where the utility may be a combination of both average latency and the
variance in the latency). Clearly, the restart strategy will always provide a
higher utility than no restart strategy, i.e. $U_{restart}(f)\geq
U_{norestart}(f)$. However, the threshold utility remains the same. Thus, as
Figure 3 illustrates for the case in which $\sigma=0$, the equilibrium point
is located at the intersection marked 1 when no agent is using the restart
strategy, and by point 2 when all agents use the restart strategy. In the
first case, a single agent who is using the restart strategy will benefit from
a much larger utility (point 1'). However, when most agents do so, this
advantage is lost, and the converse is true: if a few agents are opting to not
use the restart strategy, they will experience a utility below threshold
(point 2') and will cooperate (by not using bandwidth greedily). There are two
other interesting conclusions to the analysis above if $\sigma>0$. First, if
$f_{equ}>0.5$ the perceived equilibrium utility will be above threshold.
Second, the average perceived utility will \textit{increase} when all agents
use the restart strategy. However, this latter effect will be small if the
change in $f_{equ}$ is small. Both these results are confirmed by the simulations.

To summarize, the utility perceived by the agents using the dominant strategy
will not decrease as the strategy switches from no restart to restart.
However, if a small number of agents decide not to switch to the restart
strategy, they will obtain a lower utility than before. Thus, every agent is
always better off using the restart strategy, even if everyone does so. There
is no dilemma. However, such an argument does not model the effect of dynamics
and fluctuations in a large agent population, which we now study.

\section{Results}

\begin{figure}[tbh]
\begin{center}
\includegraphics[scale=0.5]{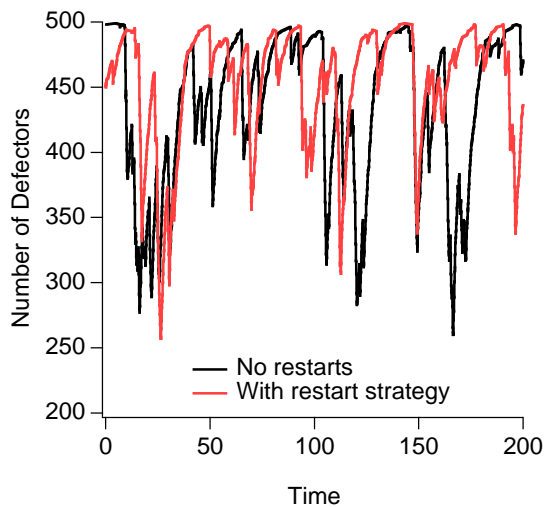}
\end{center}
\caption{Time trace of the number of defectors, both with and without use of
the restart strategy.}%
\end{figure}

A typical trace of the number of defectors (network users) as a function of
time is shown in Figure 4, with and without use of the restart strategy. The
simulations were performed with 500 agents, a threshold latency $\lambda_{c} $
of 0.05 and a variance in the noise of 0.01 with a correlation time of 1. The
histogram was collected with a relaxation time of 10000 (in units of
$\alpha^{-1}$). While the differences between the two traces in Fig. 4 is
difficult to see, the average number of defectors and the amplitude of the
fluctuations were both larger when agents made use of the restart strategy, as expected.

\begin{figure}[ptb]
\begin{center}
\includegraphics[scale=0.6]{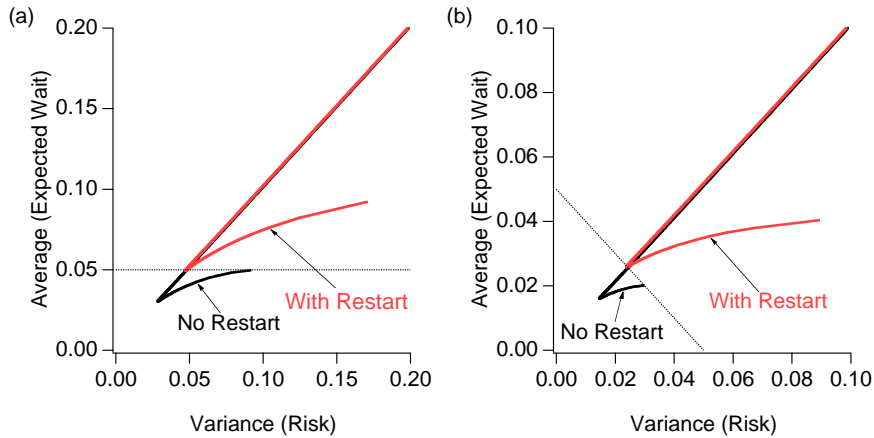}
\end{center}
\caption{Average wait vs variance (risk) as a function of the optimum restart
time when all agents do and do not make use the optimum restart strategy. The
dotted line indicates the threshold indifference curve. (a) Agents are
indifferent to risk. (b) Expected return and risk contribute equally to the
utility function.}%
\end{figure}

The two portfolio curves that result from the simulation in Figure 4 are shown
in Figure 5. The dotted line indicates the threshold latency: when agents do
not use the restart strategy, the end point of the curve (corresponding to an
infinitely long restart time) is expected to fall on the dotted line. In (a),
when agents do use the restart strategy, the waiting time at the cusp equals
the threshold latency. As expected, the optimum point (cusp) shifts to higher
average and variance as more agents make use of the restart strategy.
Similarly, a single agent who does not make use of the restart strategy while
every body else does will experience larger waiting times and larger risk, on
average, than he would have experienced with nobody using the restart
strategy. Note that the section of the portfolio curves corresponding to very
short restart times falls on the $y=x$ line -- this can be verified by
expanding the expressions for $\langle t\rangle$, $\langle t^{2}\rangle$ and
$p(t)$ for small restart times $\tau$.

\begin{figure}[tbh]
\begin{center}
\includegraphics[scale=0.5]{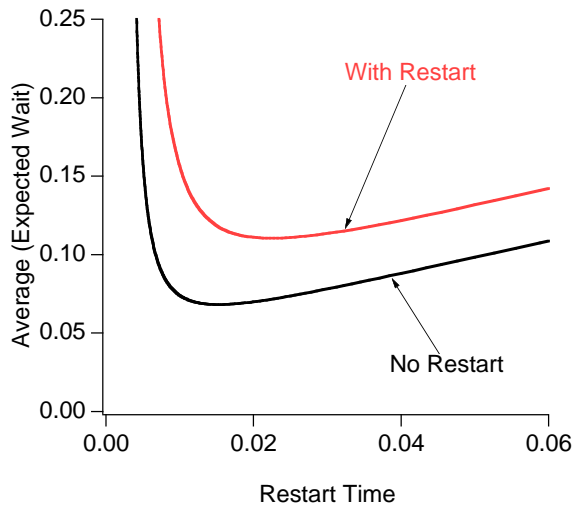}
\end{center}
\caption{Average waiting time (latency) as a function of the restart time,
with and without use of the restart strategy.}%
\end{figure}

Figure 6 shows the behavior of the expected waiting time as a function of the
restart time. As more agents make use of the restart strategy, the optimum
restart time (corresponding to the minimum expected waiting time) shifts to
larger values.

We performed an extensive set of simulations to check these results as the
parameters of the model were varied. Very short correlation times washed out
the interactions between different agents, effectively placing them on
different networks -- the restart strategy had no effect. As the correlation
time was made larger, so that a large fraction of the agents effectively
updated their decisions simultaneously, the restart strategy became more
effective. In fact, other than causing a shift in the equilibrium latencies,
noise amplitude played a relatively minor role compared to the fluctuations
due to the interaction of agents (modelled by the correlations in the noise).

\begin{figure}[ptb]
\begin{center}
\includegraphics[scale=0.6]{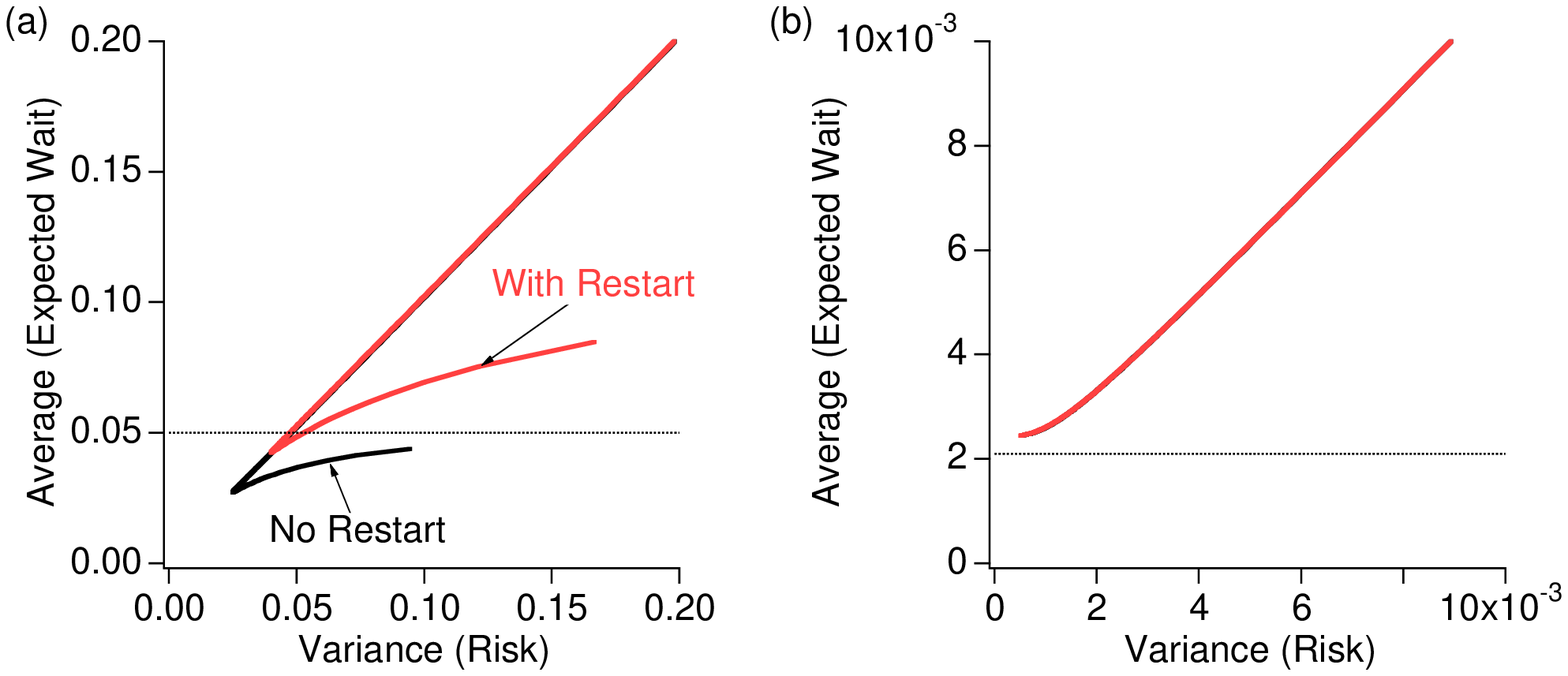}
\end{center}
\caption{Portfolio curves for the same parameters as in Figure 5, except that
$\sigma= 0.2$. (a) The threshold latency was $\lambda_{c} = 0.05$. This was
high enough that more than half the agents defected, on average. Thus, the
equilibrium latencies were lower than the threshold latency, both with a
without use of the restart strategy, as predicted by the mean field theory.
(b) The threshold latency was very low, $\lambda_{c} = 0.0021$, so that a very
small number of agents defected. Note that the equilibrium point occured at
average latencies higher than the threshold. Also note that for there
parameters, use of the restart strategy or not does not affect the perceived
latencies -- both portfolio curves overlap.}%
\end{figure}

To first order, the amplitude of the noise fluctuations had a relatively minor
effect. However, Figure 7 illustrates one of the predictions made by the mean
field theory illustrated in Figure 3: a large variance will lead to an
increased utility (lower latency) if $f_{equ} > 0.5$ but to a decreased
utility (higher latency) if $f_{equ} < 0.5$.

\begin{figure}[tbh]
\begin{center}
\includegraphics[scale=0.5]{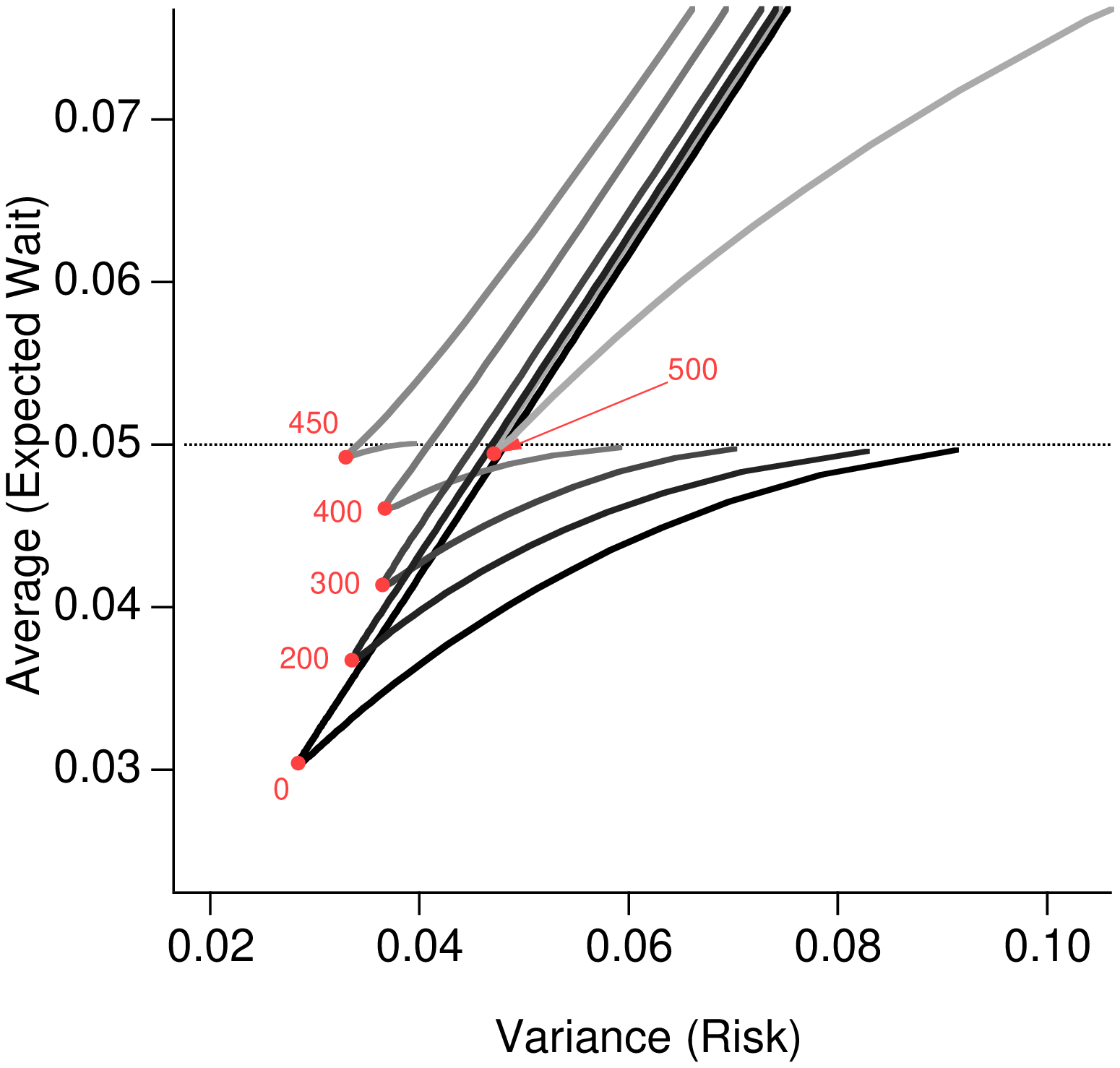}
\end{center}
\caption{Portfolio curves as the number of agents using the restart strategy
changes from 0 (no agent) to 500 (all agents).}%
\end{figure}

Figure 8 illustrates the portfolios that result from a mix of agents using and
not using the restart strategy. As the number of agents aware of the restart
strategy increases, the optimum point shifts to lower variance but higher mean
latency, until the mean latency at optimum restart equals the threshold. Until
that point, the variance for agents not using the restart strategy also
decreases. Once the number of restart agents becomes too large, the network is
too congested and the remaining agents defect.

\section{Conclusion}

In this paper we have shown that when every agent uses a portfolio strategy in
order to deal with Internet congestion, there is a range of parameters such
that a) a portfolio exists and b) all agents are better off using it than not.
Even when all agents use the optimum restart strategy, the situation is no
worse than the situation in which no one uses the restart strategy. This
solution was obtained both by using mean field arguments and computer
simulations that took into account the dynamics of agent decisions and
discounted past information on network conditions. These solutions, which show
the existence of a possibly noisy fixed point, illustrate the power of
simulations in understanding multiagent dynamics under varying dynamical constraints.

In economics, the existence of a unique and stable equilibrium is usually
considered without taking dynamics into account. Using a physics metaphor,
this is equivalent to overdamped or viscous dynamics in which velocity (rate
of change in time) does not matter. And yet, we know of a number of economic
systems which are continuously changing and sometimes are unable to reach an
equilibrium. Agent based simulations \cite{kephart}\cite{epstein}, combined
with simple analytic models, are powerful tools to study dynamic effects,
since new assumptions can be implemented and tested rapidly, and be compared
to mean field theories in order to verify their applicability.

SMM was supported by the John and Fannie Hertz Foundation.

\appendix

\section{Model Parameters and Pseudocode}

\subsection{Parameters}

\begin{itemize}
\item  Number of agents $N$

\item  Number of agents using restart strategy $N_{R}$

\item  Histogram relaxation rate $\tau_{H}$

\item  Noise correlation time $\tau$

\item  Threshold latency $\lambda_{c}$

\item  Noise variance $\sigma$ (noise mean is 1.0)
\end{itemize}

The reevaluation rate $\alpha$ for a single agent sets the time scale for the
model. We also need a function to map the number of agents using the network
into an average expected latency. We use the functional dependence expected
for a M/M/1 queue with a capacity larger than the number of users.

\subsection{Pseudocode}

\noindent REPEAT \newline \indent Pick a time step $\Delta t$ \newline
\indent\indent exponential random deviate with mean $\frac{1}{N \alpha}$ \newline 

\indent Update running user histogram \newline \indent\indent Discount
histogram using $\tau_{H}$ \newline \indent\indent Add $\Delta t$ to
appropriate bin \newline 

\indent Pick an agent randomly\newline 

\indent Compute the perceived average latency $\lambda$ \newline
\indent\indent based on the running histogram and the \newline \indent\indent
chosen agent's strategy \newline 

\indent Multiply $\lambda$ by $1 + N(0, \sigma^{2})$ \newline \indent
\indent$N(m, \sigma^{2})$ is a normal distribution with mean $\mu$ and
variance $\sigma^{2}$ \newline 

\indent IF ($\lambda< \lambda_{c}$) \newline \indent\indent Agent does use the
network (defect) \newline \indent ELSE \newline \indent\indent Agent does not
use the network (cooperate) \newline 

\noindent END REPEAT \newline 

\begin{thebibliography}{99}
\bibitem{congestion}Bernardo A. Huberman and Rajan M. Lukose. \textit{Social
Dilemmas and Internet Congestion}. Science, 277:535-537 (1997).

\bibitem{lukose}Rajan M. Lukose and Bernardo A. Huberman. \textit{A
Methodology for Managing Risk in Electronic Transactions over the Internet}.
3rd International Conference on Computational Economics, Stanford University.
June 30 1997.

\bibitem{dilemmas}Garret Hardin, \textit{The Tragedy of the Commons}, Science,
162, 1243-1248.

\bibitem{dynamics}Natalie S. Glance and Bernardo A. Huberman, \textit{Dynamics
of Social Dilemmas}, Scientific American, 76-81 (1994).

\bibitem{Huberman1997}Bernardo A. Huberman, Rajan M. Lukose and Tad Hogg.
\textit{An economics approach to hard computational problems}. Science,
275:51-54 (1997).

\bibitem{hardin}Russell Hardin, Collective Action, Johns Hopkins University
Press (1982).

\bibitem{Sharpe1970}William F. Sharpe. \textit{Portfolio Theory and Capital
Markets}. McGraw-Hill Inc., New York, 1970.

\bibitem{Huberman1988}Bernardo A. Huberman and Tad Hogg. \textit{The Behavior
of Computational Ecologies}. in The Ecology of Computation, B. A. Huberman,
Ed. (Elsevier Science Publishers B.V. (North-Holland), 1988), pp. 77-115.

\bibitem{beliefs}Bernardo A. Huberman and Natalie S. Glance. \textit{Beliefs
and Cooperation}, in Modelling Rational and Moral Agents, P. Danielson, Ed.
(Cognitive Science Series, Oxford Univ. Press, New York, 1997), pp. 211-236.

\bibitem{pnas}Bernardo A. Huberman and Natalie S. Glance. \textit{Evolutionary
Games and Computer Simulations}. Proc. Natl. Acad. Sciences (USA), 90,
7716-7718 (1993).

\bibitem{MM1queue}Sheldon M. Ross. \textit{Stochastic Processes}. John Wiley
\& Sons, Inc., New York, 1996.

\bibitem{kephart}Jeffrey O. Kephart, Tad Hogg and Bernardo A. Huberman,
\textit{Dynamics of Computational Ecosystems}, Phys. Rev A40, 404-421 (1989).

\bibitem{epstein}Growing Artificial Societies : Social Science from the Bottom
Up by Joshua M. Epstein, and Robert L. Axtell (1997).
\end{thebibliography}
\end{document}